\pdfoutput=1
\documentclass[a4paper,11pt]{article}
\usepackage[utf8]{inputenc}
\usepackage{jcappub}
\usepackage[utf8]{inputenc}
\usepackage{amsmath,amssymb}
\usepackage{graphicx}
\usepackage{multicol}
\usepackage{hyperref}
\usepackage{physics}
\usepackage{lipsum}

\usepackage{pgfplots}
\usepgfplotslibrary{fillbetween}
\usetikzlibrary{patterns}
\pgfplotsset{compat=1.8}
\newcommand{\defpicheight}{5cm}

\newcommand{\be}{\begin{equation}} 
\newcommand{\ee}{\end{equation}}
\newcommand{\bea}{\begin{equation}\begin{aligned}} 
\newcommand{\eea}{\end{aligned}\end{equation}}
\newcommand{\ba}{\begin{eqnarray}}
\newcommand{\ea}{\end{eqnarray}}

\title{Gravitational dark matter production in Palatini preheating}
\author{Alexandros Karam,}
\author{Martti Raidal,}
\author{and Eemeli Tomberg}
\emailAdd{alexandros.karam@kbfi.ee} 
\emailAdd{martti.raidal@cern.ch} 
\emailAdd{eemeli.tomberg@kbfi.ee}
\vspace{5cm}
\affiliation{Laboratory of High Energy and Computational Physics, National Institute of Chemical Physics and Biophysics, R\"avala pst.~10, 10143 Tallinn, Estonia
}

\abstract{We study preheating in plateau inflation in the Palatini formulation of general relativity, in a special case that resembles Higgs inflation. It was previously shown that the oscillating inflaton field returns to the plateau repeatedly in this model, and this leads to tachyonic production of inflaton particles. We show that a minimally coupled spectator scalar field can be produced even more efficiently by a similar mechanism. The mechanism is purely gravitational, and the scalar field mass can be of order $10^{13}$ GeV, larger than the Hubble scale by many orders of magnitude, making this a candidate for superheavy dark matter.}


\begin{document}

\maketitle

\section{Introduction} \label{sec:intro}

The nature of dark matter (DM) is one of the outstanding puzzles of contemporary high energy physics. It comprises around $84\%$ of the matter content of the
Universe~\cite{Aghanim:2018eyx} but, although it is believed to be of particulate nature, the existence of DM has been established only through its gravitational interactions. The absence of convincing DM signals in terrestrial experiments indicates that DM has negligible or non-existent interactions with the Standard Model (SM) particles. An appealing idea is that DM is gravitationally produced during inflation or shortly after its end~\cite{Ford:1986sy}. Various ways to gravitationally produce DM particles with inflation have been investigated in the past years \cite{Markkanen:2015xuw, Fairbairn:2018bsw, Bernal:2018hjm, Hashiba:2018tbu, Velazquez:2019mpj, Herring:2019hbe,  Herring:2020cah, Ahmed:2020fhc, Laulumaa:2020pqi}, also in the case where DM is superheavy \cite{Chung:1998zb, Kuzmin:1998kk, Chung:2001cb, Kannike:2016jfs, Ema:2018ucl, Li:2019ves, Ema:2019yrd, Babichev:2020xeg, Babichev:2020yeo}.

Regarding inflation, the latest data from the Planck satellite~\cite{Akrami2018} have severely constrained the allowed values for the scalar spectral index $n_s$ and the tensor-to-scalar ratio $r$. As a consequence, simple inflationary models with monomial potentials have been ruled out. On the other hand, models where the inflaton $\phi$ is non-minimally coupled to gravity through a term of the form $\xi \phi^2 R$ (such as Higgs inflation~\cite{Bezrukov2008, Bezrukov2009a, Rubio:2018ogq}) seem to be favoured, where $\xi$ is the non-minimal coupling and $R$ the Ricci scalar. 
If the Higgs boson is non-minimally coupled to gravity, predictions for the  Higgs inflation parameters are different in the metric and Palatini formulations of the theory~\cite{Bauer:2010jg, Rasanen2017, Rasanen2018, Jinno2020}\footnote{In the Palatini or first-order variational approach~\cite{Palatini1919, Ferraris1982, Exirifard2008}, the connection and metric are assumed to be independent degrees of freedom and one has to vary the action with respect to both of them. In contrast, in the metric or second-order formalism the connection is the Levi-Civita and the action is only varied with respect to the metric. Note that even though the two variational approaches result in the same equations of motion for an action which is minimally coupled and linear in $R$, the same does not hold for more complicated actions (see e.g.~\cite{Exirifard2008, Bauer2008, Bauer2011, Tamanini2011, Bauer:2010jg, Olmo2011, Rasanen2017, Tenkanen:2017jih, Racioppi2017, Markkanen:2017tun, Jaerv2018, Fu:2017iqg, Racioppi2018, Carrilho:2018ffi, Kozak:2018vlp, Bombacigno2019, Enckell2019, Rasanen2019, Antoniadis2018, Rasanen2018, Almeida2019,  Antoniadis2019, Shimada2019, Takahashi2019, Jinno2019, Tenkanen2019, Edery2019, Rubio:2019ypq, Jinno2020, Aoki2019, Giovannini2019, Tenkanen2019a, Bostan2019a, Bostan2019, Tenkanen2020, Gialamas2020, Racioppi2019, Antoniadis2019a, Tenkanen2020a, Tenkanen2020b, Shaposhnikov:2020fdv, LloydStubbs2020, Antoniadis2020, Borowiec:2020lfx, Ghilencea2020, Das:2020kff, Jarv:2020qqm, Gialamas:2020snr}).}. A notable example of a difference between the two formulations is the tensor-to-scalar ratio $r$, which is predicted to be much smaller in the Palatini formulation.
This allows, in principle, to use the measurements of inflationary parameters to distinguish between different formulations of gravity.   

Similarly, the cosmology can be very different in the two formulations of Higgs inflation. It has been demonstrated~\cite{Fu:2017iqg, Rubio:2019ypq} that the preheating stage of Higgs inflation in the two theories is dominated by different mechanisms and proceeds via different channels. While in the metric formulation preheating occurs predominantly via the parametric resonance~\cite{PhysRevD.42.2491, Kofman:1994rk, Shtanov:1994ce, Kofman:1997yn} into the vector boson channels~\cite{Bezrukov:2008ut, GarciaBellido:2008ab, DeCross:2015uza, Repond:2016sol, Ema:2016dny, DeCross:2016fdz, DeCross:2016cbs, Sfakianakis:2018lzf}, in the Palatini version of the theory the dominant mechanism is tachyonic resonance~\cite{Felder:2000hj, Felder:2001kt} into the Higgs boson itself. Therefore the preheating mechanism is much more efficient in the Palatini formalism~\cite{Rubio:2019ypq}.

In this paper, we extend the analysis of \cite{Rubio:2019ypq} by adding to the model a free scalar field, minimally coupled to gravity. We show that even in such a minimal case, the scalar field may get highly excited during preheating, and even surpass inflaton particle production as the leading preheating channel. This is true even in a supermassive case where the scalar field mass exceeds the Hubble scale of inflation by many orders of magnitude. This is the consequence of a tachyonic instability, apparent in the Einstein frame, similar to the tachyonic instability of the inflaton field itself.

The paper is organized as follows. In Sec.~\ref{sec:model} we set up the model and derive constraints on the parameters through the inflationary observables. Then, in Sec.~\ref{sec:particles} we study the production of inflaton and DM particles during the preheating stage. In Sec.~\ref{sec:results} we present our results and derive a bound on the DM mass. Finally, we conlcude in Sec.~\ref{sec:conclusions}.

\section{Model} \label{sec:model}

\subsection{Action}
Let us consider the following action for the inflaton $\phi$ and a spectator scalar field $\chi$, adopting the Palatini formulation of GR (with reduced Planck mass $M_{\rm Pl} \equiv 1$):
\begin{equation} \label{eq:SJordan}
    \begin{split}
	S = \int \dd^4 x \sqrt{-g} \bigg[ & \frac{1}{2}\qty(1 + \xi \phi^2 )g^{\mu\nu}R_{\mu\nu}\qty(\Gamma) - \frac{1}{2}g^{\mu\nu}\partial_\mu\phi\partial_\nu\phi - \frac{\lambda}{4}\phi^4 \\	&- \frac{1}{2}g^{\mu\nu}\partial_\mu\chi\partial_\nu\chi - \frac{1}{2}m_\chi^2 \chi^2 - \frac{\alpha}{2} \chi^2 \phi^2 \bigg] \, .
    \end{split}
\end{equation}
The inflationary dynamics of this model have been studied in the context of Higgs inflation \cite{Bauer2008}, though we do not assume $\phi$ to necessarily be the Higgs; preheating without $\chi$ was studied in \cite{Rubio:2019ypq}. In the Jordan frame action \eqref{eq:SJordan} we have considered a minimal setup for $\chi$: it is a free scalar field\footnote{The $\alpha$-coupling between $\phi$ and $\chi$ is included to prevent the production of $\chi$-particles during inflation; see section \ref{sec:dm_paticles}. Very small values of $\alpha$ are sufficient for this purpose, and it does not affect subsequent dynamics.}. The field $\chi$ is minimally coupled to gravity, in contrast to the inflaton, and unlike in most models of gravitational particle production. During preheating, $\chi$-particles may still be produced since the oscillatory behaviour of $\phi$ induces oscillations in the metric which couples to $\chi$. This type of `gravitational preheating' has been studied before, but not in the context of Palatini gravity. The preheating production of $\phi$-particles is rapid in the Palatini formulation due to a tachyonic instability \cite{Rubio:2019ypq}, and we will see similar behaviour in the spectator sector.

\begin{figure}[t]
\begin{center}
\includegraphics[width=0.7\textwidth]{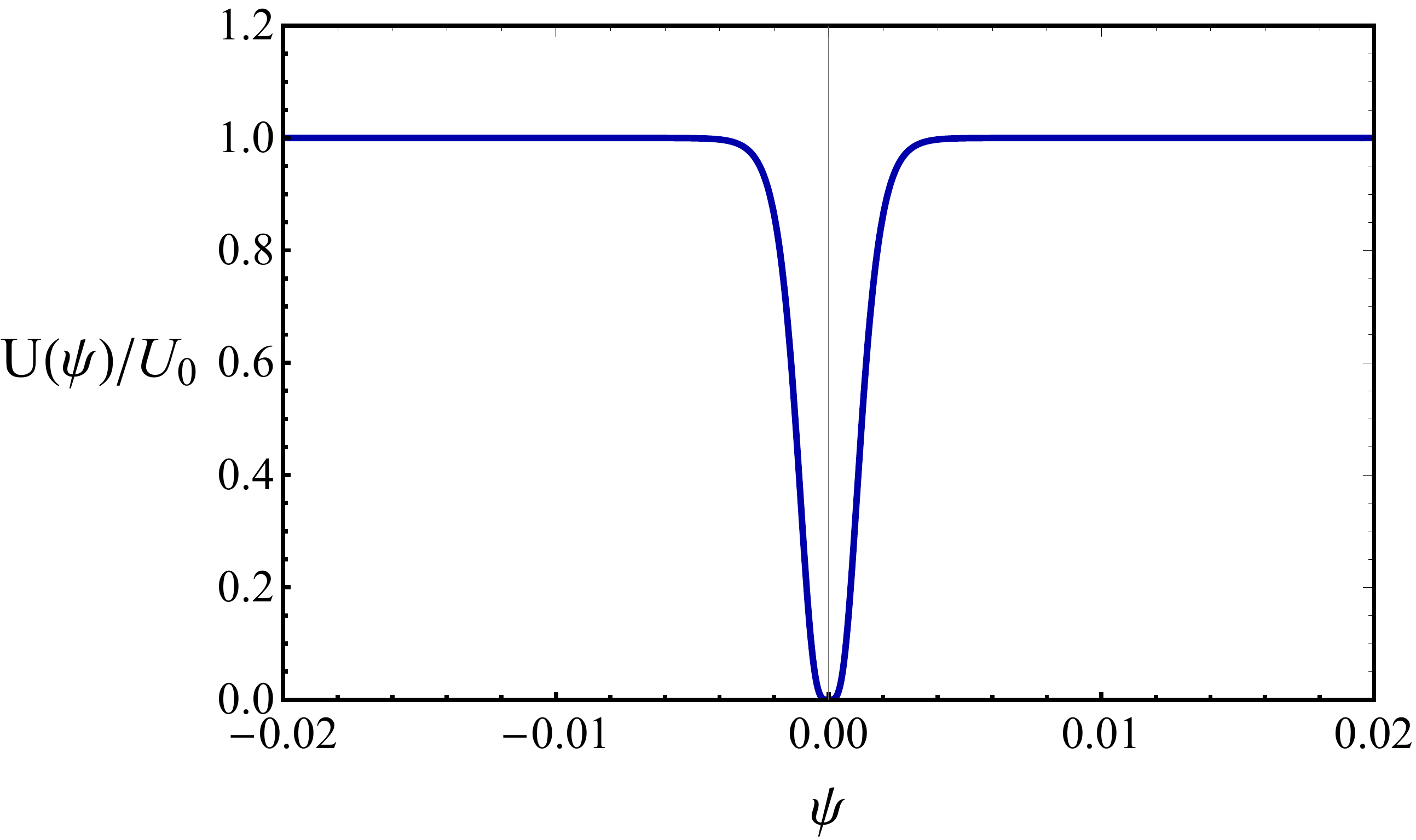}
\caption{\sf Inflaton potential for $\xi = 10^6$ and $\lambda = 10^{-2}$, with $U_0 = \frac{\lambda}{4 \xi^2}$.
}
\label{fig:potential}
\end{center}
\end{figure}

We eliminate the non-minimal coupling term by employing a Weyl transformation of the metric of the form
\begin{equation} \label{eq:metric_recaling}
	g_{\mu\nu} \rightarrow \Omega^{-2}(\phi)g_{\mu\nu} \, ,
\end{equation}
with the conformal factor given by
\begin{equation} \label{eq:Omega}
	\Omega = \sqrt{1 + \xi \phi^2} \, .
\end{equation}
Note that in Palatini gravity, $R_{\mu\nu}$ does not change in the conformal transformation, since it depends only on the connection $\Gamma$ and not on the metric. The action becomes
\begin{equation} \label{eq:SEinstein1}
\begin{split}
	S_E = \int \dd^4 x \sqrt{-g} \bigg[& \frac{1}{2}g^{\mu\nu}R_{\mu\nu}\qty(\Gamma) - \frac{1}{2\Omega^2(\phi)}g^{\mu\nu}\partial_\mu\phi\partial_\nu\phi - \frac{\lambda \phi^4}{4\Omega^4(\phi)} \\
	&- \frac{1}{2\Omega^2(\phi)}g^{\mu\nu}\partial_\mu\chi\partial_\nu\chi - \frac{1}{2\Omega^4(\phi)}m_\chi^2 \chi^2 - \frac{\alpha}{2\Omega^4(\phi)} \chi^2 \phi^2 \bigg] \, .
\end{split}
\end{equation}
We can make the inflaton kinetic term in~\eqref{eq:SEinstein1} canonical through a field redefinition of the form
\begin{equation} \label{eq:inflaton_redef}
	\frac{d\psi}{d\phi} = \Omega^{-1}(\phi) = \frac{1}{\sqrt{1+\xi\phi^2}} \,,
\end{equation}
which can be easily integrated to give 
\be 
\psi = \frac{1}{\sqrt{\xi}} \sinh^{-1} \left( \sqrt{\xi} \phi \right)
\quad \iff \quad
\phi = \frac{1}{\sqrt{\xi}} \sinh(\sqrt{\xi}\psi) \, .
\ee
Then, the Einstein frame action for the inflaton reads
\begin{equation} \label{eq:SE_infl}
	S_\psi = \int \dd^4 x \sqrt{-g} \qty[ \frac{1}{2}g^{\mu\nu}R_{\mu\nu}\qty(\Gamma) - \frac{1}{2}g^{\mu\nu}\partial_\mu\psi\partial_\nu\psi - U(\psi) ] \, ,
\end{equation}
with the potential given by
\be \label{eq:U}
U(\psi) = \frac{\lambda}{4\xi^2} \tanh^4\qty(\sqrt{\xi}\psi) \, ,
\ee
which is shown in Fig. \ref{fig:potential}. At large field values the potential is asymptotically flat and allows for inflation in accordance with the observations.

\subsection{Background dynamics} \label{sec:background}
For an FRW background, $\psi$ follows the Friedmann and Klein-Gordon equations
\begin{equation} \label{eq:FRW}
	3H^2 = \frac{1}{2}\dot{\psi}^2 + U \, , \qquad \qquad \ddot{\psi} + 3H\dot{\psi} + U' = 0 \, .
\end{equation}
The potential \eqref{eq:U} has a plateau in the large field regime where slow-roll can occur. There, \eqref{eq:FRW} become
\begin{equation} \label{eq:SR}
	3H^2 \approx U \, , \qquad \qquad 3H\dot{\psi} + U' \approx 0 \, .
\end{equation}
The duration of inflation is quantified by the number of $e$-folds which, in the slow-roll approximation, read
\begin{equation} \label{eq:efolds}
    N = \int_{\psi_{\rm end}}^{\psi} \frac{U(\tilde{\psi})}{U'(\tilde{\psi})} \dd \tilde{\psi} \approx \frac{1}{16 \xi} \cosh \left( 2 \sqrt{\xi} \psi \right) \approx \frac{\phi^2}{8} \, .    
\end{equation}
Validity of the slow-roll approximation is measured in terms of the slow-roll parameters
\begin{equation}
    \epsilon_H \equiv \frac{\dot{\psi}^2}{2H^2} \approx \frac{1}{8\xi N^2} \, , \qquad \qquad
    \eta_H \equiv -\frac{\ddot{\psi}}{H\dot{\psi}} \approx -\frac{1}{N} \, ,
\end{equation}
where the approximations are valid in slow-roll and show that $\epsilon_H$ and $\eta_H$ are small there.

The CMB observables are
\begin{equation} \label{eq:cmb_observables}
    A_s \approx \frac{1}{24 \pi^2} \frac{2 \lambda N_*^2}{\xi} \,, \qquad \quad
n_s \approx 1 - \frac{2}{N_*} \,, \qquad \quad
r \approx \frac{2}{\xi N_*^2} \,,
\end{equation}
where $N_* \approx 50$ at the pivot scale. The prediction for the spectral index, $n_s \approx 0.96$, is compatible with observations~\cite{Akrami:2018odb}. From the measured value of the amplitude of the scalar power spectrum $A_s = 2.1 \times 10^{-9}$ we obtain the relation
\begin{equation} \label{eq:pert_norm_couplings}
	\xi \approx 3.8 \times 10^6\, N_*^2\, \lambda \, .
\end{equation}
This implies a large value for $\xi$, which suppresses the tensor-to-scalar ratio $r$.

After slow-roll ends, the inflaton starts to oscillate around the potential minimum. As shown in \cite{Rubio:2019ypq}, the oscillation amplitude stays almost constant for a long time with ${\sqrt{\xi}\psi_\mathrm{max} \sim \mathcal{O}(1)}$, so that the inflaton returns repeatedly to the plateau. This is due to the specific form of the potential \eqref{eq:U} and the large value of $\xi$, and sets the model apart from most models of inflation. This behaviour has important implications for preheating, as we will see in  section \ref{sec:particles}.

During the oscillating phase, the friction term in the Klein-Gordon equation is negligible, so that \eqref{eq:FRW} become
\begin{equation} \label{eq:oscillation_approx}
	\ddot{\psi} + U' \approx 0 \, , \qquad
	3H^2 = \frac{1}{2}\dot{\psi}^2 + U \approx \text{const.} \approx \frac{\lambda}{4\xi^2} \, .
\end{equation}
From these and the potential \eqref{eq:U}, we can solve
\begin{equation} \label{eq:osc_vars}
	\ddot{\psi} \approx -\frac{\lambda}{\xi^{3/2}}\sech^2(\sqrt{\xi}\psi)\tanh^3(\sqrt{\xi}\psi) \, , \qquad
	\dot{\psi} \approx \sqrt{\frac{\lambda}{2\xi^2}\qty[1-\tanh^4(\sqrt{\xi}\psi)]} \, .
\end{equation}
In the next section, we will use these approximations to analyze particle production during preheating.

\section{Particle production} \label{sec:particles}

\subsection{Inflaton particles} \label{sec:inflaton_particles}
On top of the homogeneous background field $\psi(t)$, we have perturbations, here denoted by $q(x)$. Expanding \eqref{eq:SE_infl} to second order around the background, we get for them the linear-order action:
\begin{equation} \label{eq:Sq}
    S_q = \int \dd^4 x \, \sqrt{-g} \qty[ -\frac{1}{2}g^{\mu\nu}\partial_\mu q \partial_\nu q - U''(\psi)q^2 ] \, .
\end{equation}
Inflaton perturbations are coupled to metric perturbations, but in our model this can be ignored during preheating, as shown in \cite{Rubio:2019ypq}. Action \eqref{eq:Sq} then gives equations of motion for the Fourier modes
\begin{gather} \label{eq:q_eom}
	\ddot{q}_k + 3H\dot{q}_k + \left( \frac{k^2}{a^2} + U''(\psi) \right) q_k = 0 \, ,
\end{gather}
which we solve starting from the Bunch-Davies vacuum conditions during inflation \cite{Birrell:1982ix}:
\begin{equation} \label{eq:q_ini_conds}
    q_k = \frac{1}{a^{3/2}\sqrt{2k}} \, , \qquad \quad \dot{q}_k = -\frac{ik}{a}q_k \, , \qquad \quad k \gg aH \, .
\end{equation}
At the edge of the inflationary plateau, the mass-squared term $U''(\psi)$ is strongly negative. There the mode functions $q_k$ grow exponentially. The background field returns to this region repeatedly, yielding efficient tachyonic production of inflaton particles. Due to the tachyonic nature of the modes, the concept of particle number is not well-defined; there is no adiabatic vacuum with respect to which particles can be counted. Instead, we follow the evolution of the quantum expectation value of the energy density in the perturbations,
\begin{equation} \label{eq:q_energy}
	\rho_q = \int_0^{k_\mathrm{max}} \frac{\dd k\, k^2}{4\pi^2} \left[|\dot{q}_k|^2  +\left( \frac{k^2}{a^2} + U''(\psi) \right) |q_k|^2  \right] \, ,
\end{equation}
regulated with a momentum cut-off $k_\mathrm{max}$ which removes non-excited UV-modes.
Preheating is complete when $\rho_q$ becomes comparable to the background energy density. Then the inflaton condensate has fragmented into particles. As it turns out, this happens almost instantaneously, in only a few oscillations of the background field, less than one e-fold of expansion \cite{Rubio:2019ypq}.

If we assume the inflaton to be the Higgs, tachyonic preheating leads to exponential creation of Higgs excitations and the longitudinal mode of the Standard Model gauge bosons. These thermalize into SM plasma, completing reheating. If the inflaton is not the Higgs, it can still decay to SM particles through additional couplings, possibly going through the Higgs first. Alternatively, preheating can proceed through the production of spectator field particles, to be discussed in the next section. 

\subsection{Spectator field particles} \label{sec:dm_paticles}

To study the production of $\chi$-particles, we start with the $\chi$-dependent part of the action \eqref{eq:SEinstein1}:
\begin{equation} \label{eq:Schi}
	S_\chi = \int \dd^4 x \sqrt{-g} \qty[ -\frac{1}{2\Omega^2(\phi)}g^{\mu\nu}\partial_\mu \chi \partial_\nu \chi - \frac{1}{2\Omega^4(\phi)}m_\chi^2 \chi^2 - \frac{\alpha}{2\Omega^4(\phi)} \chi^2 \phi^2 ] \, .
\end{equation}
To get a canonical action, we make a field redefinition
\begin{equation} \label{eq:DM_redef}
	\sigma \equiv \Omega^{-1}(\phi) \chi \, ,
\end{equation}
and the action becomes
\begin{equation} \label{eq:SE_DM}
	S_\sigma = \int \dd^4 x \sqrt{-g} \qty[ - \frac{1}{2}g^{\mu\nu}\partial_\mu\sigma\partial_\nu\sigma - g^{\mu\nu}\frac{\sigma \partial_\mu \sigma \partial_\nu \Omega}{\Omega} - \frac{1}{2}\qty( \frac{m_\chi^2 + \alpha\phi^2}{\Omega^2} + g^{\mu\nu}\frac{\partial_\mu \Omega \partial_\nu \Omega}{\Omega^2} ) \sigma^2 ] \, .
\end{equation}
The equations of motion in Fourier space read
\begin{equation} \label{eq:sigma_Fourier_eom}
	\ddot{\sigma}_k + 3H\dot{\sigma}_k + \underbrace{\qty[\frac{k^2}{a^2} + \frac{m_\chi^2 + \alpha\phi^2}{\Omega^2} + 3H\frac{\dot{\Omega}}{\Omega} - 2\qty(\frac{\dot{\Omega}}{\Omega})^2 + \frac{\ddot{\Omega}}{\Omega}]}_{\equiv \, \omega_k^2} \sigma_k = 0 \, ,
\end{equation}
with the Bunch-Davies initial conditions
\begin{equation}
	\sigma_k = \frac{1}{a^{3/2}\sqrt{2\omega_k}} \, , \qquad \qquad \dot{\sigma}_k = -i\sqrt{\omega_k}\sigma_k \, ,
\end{equation}
set at a time before preheating when $\omega_k$ changes slowly and the mode evolves adiabatically. These results are general and apply for any conformal factor $\Omega$; a conformal transformation like ours generates an explicit coupling between $\chi$ and the inflaton field $\phi$ even if none is present in the Jordan frame.

Using results from section \ref{sec:background}, the $\Omega$-terms can be written as
\begin{equation} \label{eq:Omega_terms_SR}
	\frac{\dot{\Omega}}{\Omega} = \sqrt{2\xi\epsilon_H}\tanh\qty(\sqrt{\xi}\psi) H \, , \qquad \frac{\ddot{\Omega}}{\Omega} = \qty[2\xi\epsilon_H - \sqrt{2\xi\epsilon_H}\eta_H\tanh\qty(\sqrt{\xi}\psi)] H^2 \, .
\end{equation}
During inflation, the $\Omega$-terms are slow-roll-suppressed. However, as is well known, the friction term $3H\dot{\sigma}_k$ is important; it can be transformed into a mass term by a further field redefinition,
\begin{equation} \label{eq:theta_def}
    \theta \equiv a^3 \sigma \, ,
\end{equation}
so that, dropping the slow-roll-suppressed terms,
\begin{equation} \label{eq:theta_eom}
    \ddot{\theta}_k + \qty[\frac{k^2}{a^2} + \frac{m_\chi^2 + \alpha\phi^2}{\Omega^2} - 9H^2]\theta_k \approx 0 \, .
\end{equation}
The new $9H^2$-term gives a tachyonic mass contribution and may lead to strong amplification of super-Hubble modes already during inflation. This is especially true in our model, since the $m_\chi^2$-term is suppressed by $\Omega^2 = 8N\xi$ (see \eqref{eq:Omega} and \eqref{eq:efolds}), and is always below $9H^2$ for large enough $N$. This is undesirable, since $\chi$-perturbations produced during inflation might easily violate isocurvature bounds \cite{Chung:2004nh, Akrami:2018odb, Tenkanen:2019aij}. Fortunately, the situation can be saved by the $\alpha$-term, which for large $\phi$ behaves as $\alpha \phi^2/\Omega^2 \approx \alpha/\xi$. No particle production occurs during inflation, if
\begin{equation} \label{eq:alpha_lower_limit}
    \frac{\alpha}{\xi} > 9H^2 = \frac{3\lambda}{4\xi^2} \quad \iff \quad \alpha > \frac{3\lambda}{4\xi} \approx 4 \times 10^{-9} \, ,
\end{equation}
where we used \eqref{eq:pert_norm_couplings}. We assume this to be true for the following. Production of $\chi$-particles then takes place exclusively during preheating.

\begin{figure}
    \centering
    \includegraphics{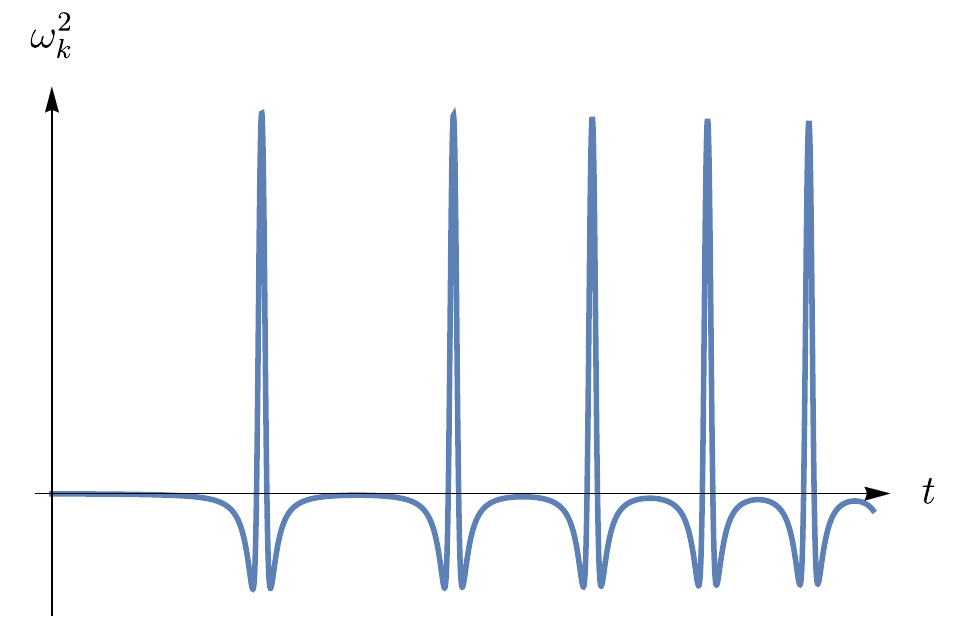}
    \caption{\sf Typical behaviour of the factor $\omega_k^2$, controlling $\chi$-production, as a function of time during preheating. There is a high positive peak when the inflaton crosses zero, but on both sides of this peak, $\omega_k^2$ is negative, signaling a tachyonic instability.}
    \label{fig:w2}
\end{figure}

During preheating, the $\Omega$-terms in \eqref{eq:sigma_Fourier_eom} are no longer suppressed. Instead, using \eqref{eq:Omega_terms_SR}, the definitions of $\epsilon_H$ and $\eta_H$, and the approximations \eqref{eq:osc_vars}, we can write $\omega_k^2$ as\footnote{As discussed in the context of inflaton perturbations above, preheating happens very fast and the universe does not expand considerably during that time, so the friction term $3H\dot{\sigma}_k$, or the corresponding tachyonic mass, are unimportant here.}
\begin{equation} \label{eq:omega2_in_theta}
	\omega_k^2 \approx \frac{k^2}{a^2} + \frac{\lambda}{\xi \cosh^2(\sqrt{\xi}\psi)} \Bigg[ \underbrace{-2 + \frac{9}{2}\sech^2(\sqrt{\xi}\psi) - 2\sech^4(\sqrt{\xi}\psi)}_{\equiv \ f(\sqrt{\xi}\psi)} + \frac{\xi m_\chi^2}{\lambda} + \frac{\alpha \sinh(\sqrt{\xi}\psi)}{\lambda}\Bigg] \, .
\end{equation}
The function $f$ reaches a minimum value of $-2$ when the inflaton is on the plateau, a negative contribution to the effective mass squared. Once again, the high oscillation amplitude of the background inflaton field leads to tachyonic particle production, this time for the $\chi$-particles, as can be seen from figure \ref{fig:w2}. Tachyonic production takes place, if
\begin{equation} \label{eq:mp}
	m_\chi \lesssim m_\mathrm{tac} \equiv \sqrt{\frac{2\lambda}{\xi}} \approx 3.5 \times 10^{13} \ \mathrm{GeV} \, ,
\end{equation}
where we again used \eqref{eq:pert_norm_couplings} and restored the units of $M_\mathrm{Pl}$. Once again, we measure particle production by computing the perturbation energy density, which now takes the form
\begin{equation} \label{eq:sigma_energy_density}
\begin{aligned}
	\rho_\sigma %\equiv -\frac{2}{\sqrt{-g}}\frac{\delta S_\sigma}{\delta g^{00}} 
	= \int_0^{k_\mathrm{max}} \frac{\dd k \, k^2}{4\pi^2} \qty[ |\dot{\sigma}_k|^2 + 2\frac{\dot{\Omega}}{\Omega}\Re \qty(\sigma_k^*\dot{\sigma}_k) + \qty( \frac{k^2}{a^2} + \frac{m_\chi^2 + \alpha\phi^2}{\Omega^2} + \frac{\dot{\Omega}^2}{\Omega^2}) |\sigma_k|{}^2 ] \, .
\end{aligned}
\end{equation}
When deriving \eqref{eq:mp}, we assumed that the $\alpha$-term is insignificant during preheating. From \eqref{eq:omega2_in_theta}, we see that this is true if $\alpha \ll \lambda$. The full limits for $\alpha$ are then
\begin{equation} \label{eq:alpha_both_limits}
	\frac{3\lambda}{4\xi} < \alpha \ll \lambda \, .
\end{equation}
For large $\xi$, this leaves a wide range of viable coupling values. It should be emphasized that $\chi$-particle production in this model is not caused by the $\alpha$-term; its purpose is merely to eliminate inflationary isocurvature perturbations and to let us focus on preheating dynamics. Indeed, large $\alpha$ suppresses the gravitational, tachyonic particle production instead of enhancing it.

We expect this non-perturbative tachyonic process to describe particle production well even if we include a quartic self-coupling for the $\chi$-field or couple it to other fields. Such couplings may play an important role later if $\chi$ is dark matter or decays into other particles, but during the violent tachyonic preheating they only give small, perturbative corrections to the dynamics.

\section{Results} \label{sec:results}

\subsection{Numerical scan} \label{sec:scan}

Our model has three free parameters: $m_\chi$, $\lambda$, and $\xi$. Out of these, $\lambda$ and $\xi$ are related by \eqref{eq:pert_norm_couplings}. We study all possibilities by varying these parameters; we only demand that $\lambda < 1$ for perturbativity (this translates into $\xi < 10^{10}$), and that $\xi \gtrsim 10^4$ (equivalently, $\lambda \gtrsim 10^{-6}$), since for smaller $\xi$-values, the inflaton oscillation amplitude decays too quickly to support tachyonic preheating. In practice, it is useful to parametrize $\xi$ and $\lambda$ through the Hubble parameter during inflation, $H=\sqrt{\lambda/(12\xi^2)}$ \eqref{eq:oscillation_approx}. The above bounds then become
\begin{equation} \label{eq:H_bounds}
    10^{8} \ \text{GeV} < H < 10^{11} \ \text{GeV} \, .    
\end{equation}
The exact value of the parameter $\alpha$ is not important, as long as it is within the bounds \eqref{eq:alpha_both_limits}.

For each $(H,m_\chi)$ pair, we solve numerically the background equations \eqref{eq:FRW} and the perturbation equations \eqref{eq:q_eom} and \eqref{eq:sigma_Fourier_eom} over a range of relevant $k$-values. We then compute the perturbation energy densities \eqref{eq:q_energy} and \eqref{eq:sigma_energy_density} as functions of time and compare them to the background energy density $\rho_\mathrm{bg}=3H^2$.

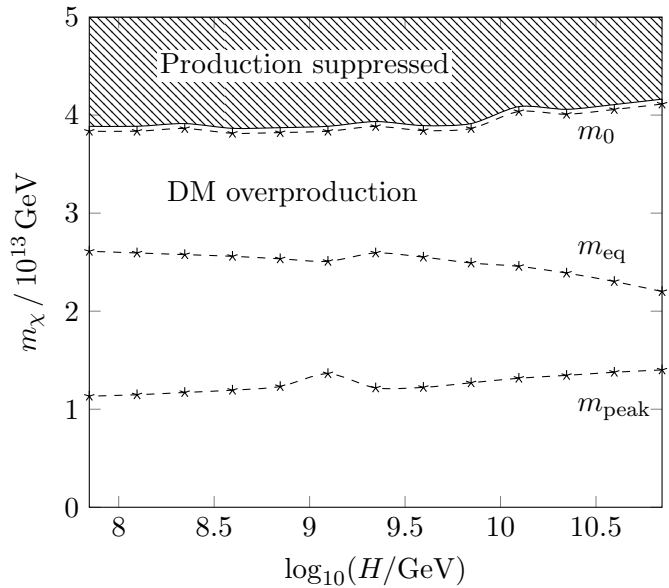
\begin{figure}
    \centering
\begin{tikzpicture}
	\begin{axis}[
		height=1.3*\defpicheight,
		scale only axis=true,
		domain=7.84597:10.846,
		xlabel=$\log_{10}(H/\mathrm{GeV})$,
		ylabel=$m_\chi \, / \, 10^{13} \, \mathrm{GeV}$,
		ymin=0,
		ymax=5,
		enlargelimits=false
		]
		\addplot[mark=none, smooth, name path=mzero] coordinates 
{(10.846, 4.16263)
(10.596, 4.10724)
(10.346, 4.05878)
(10.096, 4.08647)
(9.84597, 3.91338)
(9.59597, 3.89261)
(9.34597, 3.93415)
(9.09597, 3.88568)
(8.84597, 3.87184)
(8.59597, 3.86491)
(8.34597, 3.91338)
(8.09597, 3.88568)
(7.84597, 3.88568)};
	\addplot[mark=star, smooth, dashed, name path=mzero2] coordinates 
{(10.846, 4.16263-0.05)
(10.596, 4.10724-0.05)
(10.346, 4.05878-0.05)
(10.096, 4.08647-0.05)
(9.84597, 3.91338-0.05)
(9.59597, 3.89261-0.05)
(9.34597, 3.93415-0.05)
(9.09597, 3.88568-0.05)
(8.84597, 3.87184-0.05)
(8.59597, 3.86491-0.05)
(8.34597, 3.91338-0.05)
(8.09597, 3.88568-0.05)
(7.84597, 3.88568-0.05)};
		\addplot[mark=star, smooth, dashed, name path=meq] coordinates
{(10.846, 2.20037)
(10.596, 2.30307)
(10.346, 2.38865)
(10.096, 2.45711)
(9.84597, 2.49135)
(9.59597, 2.55125)
(9.34597, 2.59404)
(9.09597, 2.50846)
(8.84597, 2.53414)
(8.59597, 2.55981)
(8.34597, 2.57693)
(8.09597, 2.59404)
(7.84597, 2.61116)};
	\addplot[mark=star, smooth, dashed, name path=mpeak] coordinates
{(10.846, 1.40068)
(10.596, 1.37736)
(10.346, 1.34513)
(10.096, 1.31631)
(9.84597, 1.26968)
(9.59597, 1.22304)
(9.34597, 1.21754)
(9.09597, 1.36295)
(8.84597, 1.23195)
(8.59597, 1.19422)
(8.34597, 1.1709)
(8.09597, 1.14758)
(7.84597, 1.13317)};
	\addplot[mark=none, smooth, name path=topaxis] coordinates
{(7.84597, 5)
(10.846, 5)};
		\addplot[pattern=north west lines] fill between[of=mzero and topaxis];

		\node[fill=white, inner sep=1, anchor=west] at (axis cs: 8.2,4.5) {Production suppressed};
		\node[anchor=west] at (axis cs: 8.2,3.2) {DM overproduction};
		\node[anchor=west] at (axis cs: 10.35,3.8) {$m_0$};
		\node[anchor=west] at (axis cs: 10.35,2.6) {$m_\mathrm{eq}$};
		\node[anchor=west] at (axis cs: 10.35,1) {$m_\mathrm{peak}$};
        \end{axis}
\end{tikzpicture}
    \caption{\sf Production of $\chi$-particles as a function of the mass $m_\chi$ and the Hubble parameter during inflation, $H$. Tachyonic $\chi$-production is absent in the shaded region, but quickly leads to overproduction of dark matter below it. The dashed lines correspond to the masses of equation \eqref{eq:mass_values}. Note that the mass scale is much higher than the scale of the Hubble parameter.}
    \label{fig:production}
\end{figure}

The results are depicted in figure \ref{fig:production}. There, we have computed the $\chi$-energy density at the moment when the inflaton perturbations' energy density exceeds that of the background field.\footnote{We use linear perturbation theory without any backreaction on the background, so in reality our analysis is not valid all the way up to this point; however, we believe we capture the correct orders of magnitude especially for the mass values in \eqref{eq:mass_values}.} For each fixed value of $H$, we have pinpointed three important $m_\chi$-values. At $m_\mathrm{peak}$, the $\chi$-production is the most efficient; it is more efficient than the production of inflaton particles by many orders of magnitude. The $\chi$-energy density goes down for both larger and smaller masses, but we are primarily interested in superheavy DM and concentrate on masses larger than $m_\mathrm{peak}$. For a somewhat larger mass, at $m_\mathrm{eq}$, the energy densities of the inflaton and $\chi$-particles are equal at the end of preheating. For still larger masses, above the limit $m_0$, tachyonic $\chi$-production is turned off and only a negligible amount of $\chi$-particles is produced.

These mass values are almost independent of $H$ and are all close to the analytically derived tachyonicity scale $m_\mathrm{tac}$ from \eqref{eq:mp}. Approximately,
\begin{equation} \label{eq:mass_values}
\begin{gathered}
    m_\mathrm{peak} \approx 0.4 \, m_\mathrm{tac} \approx 1.3 \times 10^{13} \ \mathrm{GeV} \, , \quad
    m_\mathrm{eq} \approx 0.7 \, m_\mathrm{tac} \approx 2.4 \times 10^{13} \ \mathrm{GeV} \, , \\
    m_0 \approx 1.2 \, m_\mathrm{tac} \approx 4.0 \times 10^{13} \ \mathrm{GeV} \, .
\end{gathered}
\end{equation}
Note that these values are close to each other: the explosive tachyonic particle production is very sensitive to the value of $m_\chi$.

The produced $\chi$-particles can be much heavier than $H$, by a factor of $\sqrt{\xi} \gg 1$. They are non-relativistic from the beginning. In the next sections, we study two explicit scenarios where such superheavy $\chi$-particles play a role in explaining dark matter. In both scenarios, we take $\phi$ to be the Standard Model Higgs field, whose mass is negligible compared to the energy scales involved in preheating, as was assumed in the action \eqref{eq:SJordan}. For the SM Higgs, $\lambda \approx 0.1$, corresponding to $\xi \approx 9.5 \times 10^8$ from \eqref{eq:pert_norm_couplings} with $N_* \approx 50$.

\subsection{Example scenario: Superheavy dark matter} \label{sec:scenario1}
First, assume that $\chi$ is stable and constitutes all of dark matter with the superheavy mass $m_\chi \sim m_\mathrm{tac} \sim 10^{13}$ GeV. This dark matter is non-relativistic from the beginning, so its energy density fraction grows during radiation domination and must therefore be small initially. This is possible if $m_\chi$ is tuned close to $m_0$ from \eqref{eq:mass_values}. Preheating is then dominated by the inflaton, SM Higgs, which decays and thermalizes into a plasma of SM particles in a standard fashion. To complete the model of cosmology, baryon asymmetry can be created e.g. through leptogenesis \cite{Fukugita:1986hr, Davidson:2008bu}.

To find the required initial $\chi$-fraction, we note that during radiation domination it grows as the scale factor, and at the time of matter-radiation equality, it should be of order one. The scale factor, in turn, is inversely proportional to temperature. Thus, initially,
\begin{equation} \label{eq:rho_ratio}
    \frac{\rho_{\chi}}{\rho} = \frac{T_\mathrm{eq}}{T_\mathrm{ini}} \, ,
\end{equation}
where $\rho_\chi$ is the initial $\chi$ energy density after the near-instantaneous preheating, $\rho \sim \lambda/\xi^2$ is the initial total energy density, $T_\mathrm{ini} \sim \rho^{1/4}$ is the initial temperature, and $T_\mathrm{eq} \approx 1$~eV is the temperature at matter-radiation equality \cite{Lyth:2009zz}. For the SM Higgs with $\lambda \approx 0.1$, the ratio is $\sim 10^{-23}$.

The scenario works if the $\chi$-particles are stable. In the action \eqref{eq:SJordan}, we included the direct coupling $\alpha$ between $\chi$ and $\phi$, through which the dark matter could, in principle, decay into Standard Model particles. The decay rate from these $2 \to 2$ scatterings obeys \cite{Peskin:1995ev}
\begin{equation} \label{eq:chi_decay}
    \Gamma_{\chi \to \phi} \sim \sigma_{2\chi \to 2\phi} v n_\chi \sim \frac{\alpha^2 \rho_\chi}{m_\chi^3} \sim 10^{-18} H \frac{\alpha^2}{\lambda^{1/4}} \ll 10^{-18} H \, ,
\end{equation}
where $\sigma_{2\chi \to 2\phi}$ is the scattering cross section, $v$ is the relative velocity of the scattering $\chi$-particles, $n_\chi$ is their number density, and we used the results \eqref{eq:pert_norm_couplings}, \eqref{eq:alpha_both_limits}, and \eqref{eq:rho_ratio} with $\lambda < 1$. We have taken the non-relativistic limit for the $\chi$-particles where $v \to 0$, $n_\chi = \rho_\chi/m_\chi$, and all particles have energies of order $m_\chi$, while the $\phi$-particles (Higgs bosons) are light. Since $\Gamma_{\chi \to \phi} \ll H$, the decay is inefficient over cosmological time scales; the $\chi$-particles are practically stable. This remains to be the case afterwards, since $\Gamma_{\chi \to \phi} \propto \rho_\chi \propto a^{-3}$, and $H \propto a^{-2}$ during radiation domination and $H \propto a^{-3/2}$ during matter domination, so $\Gamma_{\chi \to \phi}$ declines faster than $H$.

\subsection{Example scenario: Superheavy portal} \label{sec:scenario1}
As a second scenario, we assume again mass of order $m_\chi \sim 10^{13}$ GeV, but this time close to $m_\mathrm{peak}$, so that the $\chi$-particles dominate over the inflaton-Higgs in preheating. To not overclose the universe, $\chi$ must be unstable. We demand a fast decay into the Higgs through the $\alpha$-coupling, $\Gamma_{\chi \to \phi} \gg H$. Mimicking \eqref{eq:chi_decay} but noting that now $\rho_\chi \approx \rho$, this condition can be written as a lower limit for $\alpha$:
\begin{equation} \label{eq:fast_decay_condition}
    \alpha \gg 0.003 \, \lambda^{1/4} \, .
\end{equation}
We also want to produce dark matter. If $\alpha$ is tuned close to the threshold \eqref{eq:fast_decay_condition}, some remnant $\chi$-particles could play this role, but a detailed calculation of the non-equilibrium dynamics is required to figure out the details.

As a simpler example, we add to the model one more scalar field $\eta$, with mass $m_\eta$ and a coupling term $\propto \beta \chi^2 \eta^2$. This new field will constitute the dark matter; we assume the direct couplings between $\eta$ and the SM to be negligible. Then $\chi$ decays into both $\phi$ and $\eta$, and afterwards, the energy density fraction in $\eta$ is $\beta^2/\alpha^2$. We assume that the $\eta$-particles don't thermalize with themselves or with the SM, due to the weakness of the interactions. As time goes on, the velocity of the $\eta$-particles is redshifted away; after the universe has expanded by a factor of $m_\chi/m_\eta$, they become non-relativistic. After this, their energy density fraction starts to increase as discussed above, and should again be order one at matter-radiation equality. Analogously to \eqref{eq:rho_ratio}, we get
\begin{equation} \label{eq:alpha_beta_ratio}
    \frac{\beta^2}{\alpha^2} = \frac{m_\chi}{m_\eta} \frac{T_\mathrm{eq}}{T_\mathrm{ini}}  \, ,
\end{equation}
which relates $\beta$ to $\alpha$. If $\alpha$, $\lambda$ and $\xi$ are fixed, this becomes a relation between $\beta$ and $m_\eta$. A wide range of masses is possible,
\begin{equation} \label{eq:eta_mass_range}
     m_\chi \frac{T_\mathrm{eq}}{T_\mathrm{ini}} <  m_\eta \leq m_\chi \, ,
\end{equation}
where the upper limit comes from the requirement of energy conservation in $\chi$-decays, and the lower limit, equivalent to $\beta < \alpha$, comes from the requirement that DM must be non-relativistic at matter-radiation equality. This dark matter may be either cold, warm, or hot: the $\eta$-particles become non-relativistic when the SM plasma has temperature $T_\mathrm{ini} m_\eta/m_\chi$, which ranges from $T_\mathrm{eq}$ to $T_\mathrm{ini}$.

What about interactions between the SM and $\eta$? These can be mediated by $\chi$, but are always loop suppressed. The $\chi$-interactions with $\phi$ and $\eta$, proportional to $\alpha$ and $\beta$, may be strong, but the direct $\phi$--$\eta$ interactions are proportional to $\alpha\beta$ and thus further suppressed, as long as both $\alpha$ and $\beta$ are small. In addition, the large mediator mass $m_\chi$ further shuts down the interactions once the universe cools down.

As example values, we take $m_\chi = 1.3 \times 10^{13}$ GeV, $\alpha = 0.01$, $\beta = 10^{-4}$, $m_\eta = 3$~keV. With the SM value $\lambda = 0.1$, these satisfy all our constraints and produce the correct dark matter density (within the accuracy of the approximations used), while keeping the interactions between the SM and $\eta$ small.

These two examples demonstrate the cosmological applications of the gravitational particle production mechanism studied in this paper. More possibilities open up with e.g. more complicated field content or lighter $\chi$-particle mass. We leave the study of such possibilities for future work.

\section{Discussion and conclusions} \label{sec:conclusions}

We studied preheating in a model analogous to Higgs inflation in the Palatini formulation of gravity. As shown in~\cite{Rubio:2019ypq}, the oscillating inflaton field returns repeatedly to the inflationary plateau, which leads to efficient tachyonic production of inflaton particles. We showed that the same happens to spectator scalar fields: they are produced rapidly through a process that can be understood as a tachyonic instability in the Einstein frame.

Scalar field particles are produced most abundantly if they have a superheavy mass around $10^{13}$ GeV. For such a mass, their production may exceed that of inflaton particles, in which case this becomes the leading preheating channel. If the particles are stable, they can constitute all of dark matter, although their mass must be finely tuned to obtain the correct abundance. If the particles are unstable, they can act as a portal through which the inflaton condensate decays into Standard Model particles and dark matter.

The scenario presented here has many interesting cosmological consequences. It places strict constraints on scalar fields in our inflationary model: even supermassive scalar fields may easily be overproduced during preheating. Our scenario is also an example of a model where supermassive DM can be produced simultaneously with a negligible tensor-to-scalar ratio $r$. In single-field inflationary models, $r$ is proportional to the inflation energy scale $H$, so small $r$ implies small $H$, which in most models sets an upper limit for the masses of produced particles: $m \lesssim H$ (though see \cite{Kuzmin:1998kk, Kannike:2016jfs, Ema:2018ucl, Ema:2019yrd, Li:2019ves, Babichev:2020xeg, Babichev:2020yeo}). However, our model allows $m \gg H$, so there is no direct connection between $m$ and $r$. Indeed, as can be seen from \eqref{eq:cmb_observables}, $\xi \gg 1$ implies a negligible $r$, but $\chi$-particles can still be produced abundantly with masses up to $10^{13}$ GeV. Similarly, a low reheating temperature is allowed, since it is controlled by $H$.

The model studied here is very minimal: the production mechanism is purely gravitational, no coupling to the inflaton is required, and the scalar field is minimally coupled to gravity. This makes the mechanism very general, and calls for caution in such models of inflation: traditional techniques of parametric resonance and perturbative decays do not describe reheating correctly, and a large mass does not automatically protect a particle against overproduction.

We expect our findings to generalize to other models of inflation where the inflaton repeatedly returns to a plateau during its oscillation. Study of such generalizations is left for future work.

%-------------------------------------------------------------------------------
\acknowledgments
%-------------------------------------------------------------------------------

This work was supported by the Estonian Research Council grants PRG803, MOBJD381 and MOBTT5
and by the EU through the European Regional Development Fund
CoE program TK133 ``The Dark Side of the Universe."

\bibliography{Palatini_DM}

\providecommand{\href}[2]{#2}\begingroup\raggedright\begin{thebibliography}{10}

\bibitem{Aghanim:2018eyx}
{\bf Planck} Collaboration, N.~Aghanim et~al., {\it {Planck 2018 results. VI.
  Cosmological parameters}},  \href{http://arxiv.org/abs/1807.06209}{{\tt
  arXiv:1807.06209}}.

\bibitem{Ford:1986sy}
L.~H. Ford, {\it {Gravitational Particle Creation and Inflation}},  {\em Phys.
  Rev.} {\bf D35} (1987) 2955.

\bibitem{Markkanen:2015xuw}
T.~Markkanen and S.~Nurmi, {\it {Dark matter from gravitational particle
  production at reheating}},  {\em JCAP} {\bf 02} (2017) 008,
  [\href{http://arxiv.org/abs/1512.07288}{{\tt arXiv:1512.07288}}].

\bibitem{Fairbairn:2018bsw}
M.~Fairbairn, K.~Kainulainen, T.~Markkanen, and S.~Nurmi, {\it {Despicable Dark
  Relics: generated by gravity with unconstrained masses}},  {\em JCAP} {\bf
  04} (2019) 005, [\href{http://arxiv.org/abs/1808.08236}{{\tt
  arXiv:1808.08236}}].

\bibitem{Bernal:2018hjm}
N.~Bernal, A.~Chatterjee, and A.~Paul, {\it {Non-thermal production of Dark
  Matter after Inflation}},  {\em JCAP} {\bf 12} (2018) 020,
  [\href{http://arxiv.org/abs/1809.02338}{{\tt arXiv:1809.02338}}].

\bibitem{Hashiba:2018tbu}
S.~Hashiba and J.~Yokoyama, {\it {Gravitational particle creation for dark
  matter and reheating}},  {\em Phys. Rev. D} {\bf 99} (2019), no.~4 043008,
  [\href{http://arxiv.org/abs/1812.10032}{{\tt arXiv:1812.10032}}].

\bibitem{Velazquez:2019mpj}
J.~A. Cembranos, L.~J. Garay, and J.~M. Sánchez~Velázquez, {\it
  {Gravitational production of scalar dark matter}},  {\em JHEP} {\bf 06}
  (2020) 084, [\href{http://arxiv.org/abs/1910.13937}{{\tt arXiv:1910.13937}}].

\bibitem{Herring:2019hbe}
N.~Herring, D.~Boyanovsky, and A.~R. Zentner, {\it {Nonadiabatic cosmological
  production of ultralight dark matter}},  {\em Phys. Rev. D} {\bf 101} (2020),
  no.~8 083516, [\href{http://arxiv.org/abs/1912.10859}{{\tt
  arXiv:1912.10859}}].

\bibitem{Herring:2020cah}
N.~Herring and D.~Boyanovsky, {\it {Gravitational production of nearly thermal
  fermionic Dark Matter}},  {\em Phys. Rev. D} {\bf 101} (2020), no.~12 123522,
  [\href{http://arxiv.org/abs/2005.00391}{{\tt arXiv:2005.00391}}].

\bibitem{Ahmed:2020fhc}
A.~Ahmed, B.~Grzadkowski, and A.~Socha, {\it {Gravitational production of
  vector dark matter}},  \href{http://arxiv.org/abs/2005.01766}{{\tt
  arXiv:2005.01766}}.

\bibitem{Laulumaa:2020pqi}
L.~Laulumaa, T.~Markkanen, and S.~Nurmi, {\it {Primordial dark matter from
  curvature induced symmetry breaking}},
  \href{http://arxiv.org/abs/2005.04061}{{\tt arXiv:2005.04061}}.

\bibitem{Chung:1998zb}
D.~J. Chung, E.~W. Kolb, and A.~Riotto, {\it {Superheavy dark matter}},  {\em
  Phys. Rev. D} {\bf 59} (1998) 023501,
  [\href{http://arxiv.org/abs/hep-ph/9802238}{{\tt hep-ph/9802238}}].

\bibitem{Kuzmin:1998kk}
V.~Kuzmin and I.~Tkachev, {\it {Matter creation via vacuum fluctuations in the
  early universe and observed ultrahigh-energy cosmic ray events}},  {\em Phys.
  Rev. D} {\bf 59} (1999) 123006,
  [\href{http://arxiv.org/abs/hep-ph/9809547}{{\tt hep-ph/9809547}}].

\bibitem{Chung:2001cb}
D.~J. Chung, P.~Crotty, E.~W. Kolb, and A.~Riotto, {\it {On the Gravitational
  Production of Superheavy Dark Matter}},  {\em Phys. Rev. D} {\bf 64} (2001)
  043503, [\href{http://arxiv.org/abs/hep-ph/0104100}{{\tt hep-ph/0104100}}].

\bibitem{Kannike:2016jfs}
K.~Kannike, A.~Racioppi, and M.~Raidal, {\it {Super-heavy dark matter --
  Towards predictive scenarios from inflation}},  {\em Nucl. Phys. B} {\bf 918}
  (2017) 162--177, [\href{http://arxiv.org/abs/1605.09378}{{\tt
  arXiv:1605.09378}}].

\bibitem{Ema:2018ucl}
Y.~Ema, K.~Nakayama, and Y.~Tang, {\it {Production of Purely Gravitational Dark
  Matter}},  {\em JHEP} {\bf 09} (2018) 135,
  [\href{http://arxiv.org/abs/1804.07471}{{\tt arXiv:1804.07471}}].

\bibitem{Li:2019ves}
L.~Li, T.~Nakama, C.~M. Sou, Y.~Wang, and S.~Zhou, {\it {Gravitational
  Production of Superheavy Dark Matter and Associated Cosmological
  Signatures}},  {\em JHEP} {\bf 07} (2019) 067,
  [\href{http://arxiv.org/abs/1903.08842}{{\tt arXiv:1903.08842}}].

\bibitem{Ema:2019yrd}
Y.~Ema, K.~Nakayama, and Y.~Tang, {\it {Production of Purely Gravitational Dark
  Matter: The Case of Fermion and Vector Boson}},  {\em JHEP} {\bf 07} (2019)
  060, [\href{http://arxiv.org/abs/1903.10973}{{\tt arXiv:1903.10973}}].

\bibitem{Babichev:2020xeg}
E.~Babichev, D.~Gorbunov, and S.~Ramazanov, {\it {Gravitational misalignment
  mechanism of Dark Matter production}},
  \href{http://arxiv.org/abs/2004.03410}{{\tt arXiv:2004.03410}}.

\bibitem{Babichev:2020yeo}
E.~Babichev, D.~Gorbunov, S.~Ramazanov, and L.~Reverberi, {\it {Gravitational
  reheating and superheavy Dark Matter creation after inflation with
  non-minimal coupling}},  \href{http://arxiv.org/abs/2006.02225}{{\tt
  arXiv:2006.02225}}.

\bibitem{Akrami2018}
{\bf Planck} Collaboration, Y.~Akrami et~al., {\it {Planck 2018 results. X.
  Constraints on inflation}},  \href{http://arxiv.org/abs/1807.06211}{{\tt
  arXiv:1807.06211}}.

\bibitem{Bezrukov2008}
F.~L. Bezrukov and M.~Shaposhnikov, {\it {The Standard Model Higgs boson as the
  inflaton}},  {\em Phys. Lett.} {\bf B659} (2008) 703--706,
  [\href{http://arxiv.org/abs/0710.3755}{{\tt arXiv:0710.3755}}].

\bibitem{Bezrukov2009a}
F.~L. Bezrukov, A.~Magnin, and M.~Shaposhnikov, {\it {Standard Model Higgs
  boson mass from inflation}},  {\em Phys. Lett.} {\bf B675} (2009) 88--92,
  [\href{http://arxiv.org/abs/0812.4950}{{\tt arXiv:0812.4950}}].

\bibitem{Rubio:2018ogq}
J.~Rubio, {\it {Higgs inflation}},  {\em Front. Astron. Space Sci.} {\bf 5}
  (2019) 50, [\href{http://arxiv.org/abs/1807.02376}{{\tt arXiv:1807.02376}}].

\bibitem{Bauer:2010jg}
F.~Bauer and D.~A. Demir, {\it {Higgs-Palatini Inflation and Unitarity}},  {\em
  Phys. Lett.} {\bf B698} (2011) 425--429,
  [\href{http://arxiv.org/abs/1012.2900}{{\tt arXiv:1012.2900}}].

\bibitem{Rasanen2017}
S.~Rasanen and P.~Wahlman, {\it {Higgs inflation with loop corrections in the
  Palatini formulation}},  {\em JCAP} {\bf 11} (2017) 047,
  [\href{http://arxiv.org/abs/1709.07853}{{\tt arXiv:1709.07853}}].

\bibitem{Rasanen2018}
S.~Rasanen, {\it {Higgs inflation in the Palatini formulation with kinetic
  terms for the metric}},  \href{http://arxiv.org/abs/1811.09514}{{\tt
  arXiv:1811.09514}}.

\bibitem{Jinno2020}
R.~Jinno, M.~Kubota, K.-y. Oda, and S.~C. Park, {\it {Higgs inflation in metric
  and Palatini formalisms: Required suppression of higher dimensional
  operators}},  {\em JCAP} {\bf 03} (2020) 063,
  [\href{http://arxiv.org/abs/1904.05699}{{\tt arXiv:1904.05699}}].

\bibitem{Palatini1919}
A.~Palatini, {\it Deduzione invariantiva delle equazioni gravitazionali dal
  principio di hamilton},  {\em Rendiconti del Circolo Matematico di Palermo
  (1884-1940)} {\bf 43} (Dec, 1919) 203--212.

\bibitem{Ferraris1982}
M.~Ferraris, M.~Francaviglia, and C.~Reina, {\it Variational formulation of
  general relativity from 1915 to 1925 ``palatini's method'' discovered by
  einstein in 1925},  {\em General Relativity and Gravitation} {\bf 14} (Mar,
  1982) 243--254.

\bibitem{Exirifard2008}
Q.~Exirifard and M.~M. Sheikh-Jabbari, {\it {Lovelock gravity at the crossroads
  of Palatini and metric formulations}},  {\em Phys. Lett.} {\bf B661} (2008)
  158--161, [\href{http://arxiv.org/abs/0705.1879}{{\tt arXiv:0705.1879}}].

\bibitem{Bauer2008}
F.~Bauer and D.~A. Demir, {\it {Inflation with Non-Minimal Coupling: Metric
  versus Palatini Formulations}},  {\em Phys. Lett.} {\bf B665} (2008)
  222--226, [\href{http://arxiv.org/abs/0803.2664}{{\tt arXiv:0803.2664}}].

\bibitem{Bauer2011}
F.~Bauer, {\it {Filtering out the cosmological constant in the Palatini
  formalism of modified gravity}},  {\em Gen. Rel. Grav.} {\bf 43} (2011)
  1733--1757, [\href{http://arxiv.org/abs/1007.2546}{{\tt arXiv:1007.2546}}].

\bibitem{Tamanini2011}
N.~Tamanini and C.~R. Contaldi, {\it {Inflationary Perturbations in Palatini
  Generalised Gravity}},  {\em Phys. Rev.} {\bf D83} (2011) 044018,
  [\href{http://arxiv.org/abs/1010.0689}{{\tt arXiv:1010.0689}}].

\bibitem{Olmo2011}
G.~J. Olmo, {\it {Palatini Approach to Modified Gravity: f(R) Theories and
  Beyond}},  {\em Int. J. Mod. Phys.} {\bf D20} (2011) 413--462,
  [\href{http://arxiv.org/abs/1101.3864}{{\tt arXiv:1101.3864}}].

\bibitem{Tenkanen:2017jih}
T.~Tenkanen, {\it {Resurrecting Quadratic Inflation with a non-minimal coupling
  to gravity}},  {\em JCAP} {\bf 12} (2017) 001,
  [\href{http://arxiv.org/abs/1710.02758}{{\tt arXiv:1710.02758}}].

\bibitem{Racioppi2017}
A.~Racioppi, {\it {Coleman-Weinberg linear inflation: metric vs. Palatini
  formulation}},  {\em JCAP} {\bf 1712} (2017), no.~12 041,
  [\href{http://arxiv.org/abs/1710.04853}{{\tt arXiv:1710.04853}}].

\bibitem{Markkanen:2017tun}
T.~Markkanen, T.~Tenkanen, V.~Vaskonen, and H.~Veermäe, {\it {Quantum
  corrections to quartic inflation with a non-minimal coupling: metric vs.
  Palatini}},  {\em JCAP} {\bf 03} (2018) 029,
  [\href{http://arxiv.org/abs/1712.04874}{{\tt arXiv:1712.04874}}].

\bibitem{Jaerv2018}
L.~Järv, A.~Racioppi, and T.~Tenkanen, {\it {Palatini side of inflationary
  attractors}},  {\em Phys. Rev. D} {\bf 97} (2018), no.~8 083513,
  [\href{http://arxiv.org/abs/1712.08471}{{\tt arXiv:1712.08471}}].

\bibitem{Fu:2017iqg}
C.~Fu, P.~Wu, and H.~Yu, {\it {Inflationary dynamics and preheating of the
  nonminimally coupled inflaton field in the metric and Palatini formalisms}},
  {\em Phys. Rev. D} {\bf 96} (2017), no.~10 103542,
  [\href{http://arxiv.org/abs/1801.04089}{{\tt arXiv:1801.04089}}].

\bibitem{Racioppi2018}
A.~Racioppi, {\it {New universal attractor in nonminimally coupled gravity:
  Linear inflation}},  {\em Phys. Rev.} {\bf D97} (2018), no.~12 123514,
  [\href{http://arxiv.org/abs/1801.08810}{{\tt arXiv:1801.08810}}].

\bibitem{Carrilho:2018ffi}
P.~Carrilho, D.~Mulryne, J.~Ronayne, and T.~Tenkanen, {\it {Attractor Behaviour
  in Multifield Inflation}},  {\em JCAP} {\bf 06} (2018) 032,
  [\href{http://arxiv.org/abs/1804.10489}{{\tt arXiv:1804.10489}}].

\bibitem{Kozak:2018vlp}
A.~Kozak and A.~Borowiec, {\it {Palatini frames in scalar–tensor theories of
  gravity}},  {\em Eur. Phys. J.} {\bf C79} (2019), no.~4 335,
  [\href{http://arxiv.org/abs/1808.05598}{{\tt arXiv:1808.05598}}].

\bibitem{Bombacigno2019}
F.~Bombacigno and G.~Montani, {\it {Big bounce cosmology for Palatini $R^2$
  gravity with a Nieh--Yan term}},  {\em Eur. Phys. J. C} {\bf 79} (2019),
  no.~5 405, [\href{http://arxiv.org/abs/1809.07563}{{\tt arXiv:1809.07563}}].

\bibitem{Enckell2019}
V.-M. Enckell, K.~Enqvist, S.~Rasanen, and L.-P. Wahlman, {\it {Inflation with
  $R^2$ term in the Palatini formalism}},  {\em JCAP} {\bf 1902} (2019) 022,
  [\href{http://arxiv.org/abs/1810.05536}{{\tt arXiv:1810.05536}}].

\bibitem{Rasanen2019}
S.~Rasanen and E.~Tomberg, {\it {Planck scale black hole dark matter from Higgs
  inflation}},  {\em JCAP} {\bf 01} (2019) 038,
  [\href{http://arxiv.org/abs/1810.12608}{{\tt arXiv:1810.12608}}].

\bibitem{Antoniadis2018}
I.~Antoniadis, A.~Karam, A.~Lykkas, and K.~Tamvakis, {\it {Palatini inflation
  in models with an $R^2$ term}},  {\em JCAP} {\bf 1811} (2018), no.~11 028,
  [\href{http://arxiv.org/abs/1810.10418}{{\tt arXiv:1810.10418}}].

\bibitem{Almeida2019}
J.~P.~B. Almeida, N.~Bernal, J.~Rubio, and T.~Tenkanen, {\it {Hidden Inflaton
  Dark Matter}},  {\em JCAP} {\bf 03} (2019) 012,
  [\href{http://arxiv.org/abs/1811.09640}{{\tt arXiv:1811.09640}}].

\bibitem{Antoniadis2019}
I.~Antoniadis, A.~Karam, A.~Lykkas, T.~Pappas, and K.~Tamvakis, {\it {Rescuing
  Quartic and Natural Inflation in the Palatini Formalism}},  {\em JCAP} {\bf
  1903} (2019), no.~03 005, [\href{http://arxiv.org/abs/1812.00847}{{\tt
  arXiv:1812.00847}}].

\bibitem{Shimada2019}
K.~Shimada, K.~Aoki, and K.-i. Maeda, {\it {Metric-affine Gravity and
  Inflation}},  {\em Phys. Rev. D} {\bf 99} (2019), no.~10 104020,
  [\href{http://arxiv.org/abs/1812.03420}{{\tt arXiv:1812.03420}}].

\bibitem{Takahashi2019}
T.~Takahashi and T.~Tenkanen, {\it {Towards distinguishing variants of
  non-minimal inflation}},  {\em JCAP} {\bf 04} (2019) 035,
  [\href{http://arxiv.org/abs/1812.08492}{{\tt arXiv:1812.08492}}].

\bibitem{Jinno2019}
R.~Jinno, K.~Kaneta, K.-y. Oda, and S.~C. Park, {\it {Hillclimbing inflation in
  metric and Palatini formulations}},  {\em Phys. Lett. B} {\bf 791} (2019)
  396--402, [\href{http://arxiv.org/abs/1812.11077}{{\tt arXiv:1812.11077}}].

\bibitem{Tenkanen2019}
T.~Tenkanen, {\it {Minimal Higgs inflation with an $R^2$ term in Palatini
  gravity}},  {\em Phys. Rev. D} {\bf 99} (2019), no.~6 063528,
  [\href{http://arxiv.org/abs/1901.01794}{{\tt arXiv:1901.01794}}].

\bibitem{Edery2019}
A.~Edery and Y.~Nakayama, {\it {Palatini formulation of pure $R^2$ gravity
  yields Einstein gravity with no massless scalar}},  {\em Phys. Rev. D} {\bf
  99} (2019), no.~12 124018, [\href{http://arxiv.org/abs/1902.07876}{{\tt
  arXiv:1902.07876}}].

\bibitem{Rubio:2019ypq}
J.~Rubio and E.~S. Tomberg, {\it {Preheating in Palatini Higgs inflation}},
  {\em JCAP} {\bf 04} (2019) 021, [\href{http://arxiv.org/abs/1902.10148}{{\tt
  arXiv:1902.10148}}].

\bibitem{Aoki2019}
K.~Aoki and K.~Shimada, {\it {Scalar-metric-affine theories: Can we get
  ghost-free theories from symmetry?}},  {\em Phys. Rev. D} {\bf 100} (2019),
  no.~4 044037, [\href{http://arxiv.org/abs/1904.10175}{{\tt
  arXiv:1904.10175}}].

\bibitem{Giovannini2019}
M.~Giovannini, {\it {Post-inflationary phases stiffer than radiation and
  Palatini formulation}},  {\em Class. Quant. Grav.} {\bf 36} (2019), no.~23
  235017, [\href{http://arxiv.org/abs/1905.06182}{{\tt arXiv:1905.06182}}].

\bibitem{Tenkanen2019a}
T.~Tenkanen and L.~Visinelli, {\it {Axion dark matter from Higgs inflation with
  an intermediate $H_*$}},  {\em JCAP} {\bf 08} (2019) 033,
  [\href{http://arxiv.org/abs/1906.11837}{{\tt arXiv:1906.11837}}].

\bibitem{Bostan2019a}
N.~Bostan, {\it {Non-minimally coupled quartic inflation with Coleman-Weinberg
  one-loop corrections in the Palatini formulation}},
  \href{http://arxiv.org/abs/1907.13235}{{\tt arXiv:1907.13235}}.

\bibitem{Bostan2019}
N.~Bostan, {\it {Quadratic, Higgs and hilltop potentials in the Palatini
  gravity}},  \href{http://arxiv.org/abs/1908.09674}{{\tt arXiv:1908.09674}}.

\bibitem{Tenkanen2020}
T.~Tenkanen, {\it {Trans-Planckian censorship, inflation, and dark matter}},
  {\em Phys. Rev. D} {\bf 101} (2020), no.~6 063517,
  [\href{http://arxiv.org/abs/1910.00521}{{\tt arXiv:1910.00521}}].

\bibitem{Gialamas2020}
I.~D. Gialamas and A.~Lahanas, {\it {Reheating in $R^2$ Palatini inflationary
  models}},  {\em Phys. Rev. D} {\bf 101} (2020), no.~8 084007,
  [\href{http://arxiv.org/abs/1911.11513}{{\tt arXiv:1911.11513}}].

\bibitem{Racioppi2019}
A.~Racioppi, {\it {Non-Minimal (Self-)Running Inflation: Metric vs. Palatini
  Formulation}},  \href{http://arxiv.org/abs/1912.10038}{{\tt
  arXiv:1912.10038}}.

\bibitem{Antoniadis2019a}
I.~Antoniadis, A.~Karam, A.~Lykkas, T.~Pappas, and K.~Tamvakis, {\it
  {Single-field inflation in models with an $R^2$ term}},  in {\em {19th
  Hellenic School and Workshops on Elementary Particle Physics and Gravity}},
  12, 2019.
\newblock \href{http://arxiv.org/abs/1912.12757}{{\tt arXiv:1912.12757}}.

\bibitem{Tenkanen2020a}
T.~Tenkanen, {\it {Tracing the high energy theory of gravity: an introduction
  to Palatini inflation}},  {\em Gen. Rel. Grav.} {\bf 52} (2020), no.~4 33,
  [\href{http://arxiv.org/abs/2001.10135}{{\tt arXiv:2001.10135}}].

\bibitem{Tenkanen2020b}
T.~Tenkanen and E.~Tomberg, {\it {Initial conditions for plateau inflation: a
  case study}},  \href{http://arxiv.org/abs/2002.02420}{{\tt
  arXiv:2002.02420}}.

\bibitem{Shaposhnikov:2020fdv}
M.~Shaposhnikov, A.~Shkerin, and S.~Zell, {\it {Quantum Effects in Palatini
  Higgs Inflation}},  \href{http://arxiv.org/abs/2002.07105}{{\tt
  arXiv:2002.07105}}.

\bibitem{LloydStubbs2020}
A.~Lloyd-Stubbs and J.~McDonald, {\it {Sub-Planckian $\phi^{2}$ Inflation in
  the Palatini Formulation of Gravity with an $R^2$ term}},
  \href{http://arxiv.org/abs/2002.08324}{{\tt arXiv:2002.08324}}.

\bibitem{Antoniadis2020}
I.~Antoniadis, A.~Lykkas, and K.~Tamvakis, {\it {Constant-roll in the
  Palatini-$R^2$ models}},  {\em JCAP} {\bf 04} (2020), no.~04 033,
  [\href{http://arxiv.org/abs/2002.12681}{{\tt arXiv:2002.12681}}].

\bibitem{Borowiec:2020lfx}
A.~Borowiec and A.~Kozak, {\it {New class of hybrid metric-Palatini
  scalar-tensor theories of gravity}},
  \href{http://arxiv.org/abs/2003.02741}{{\tt arXiv:2003.02741}}.

\bibitem{Ghilencea2020}
D.~Ghilencea, {\it {Palatini quadratic gravity: spontaneous breaking of gauged
  scale symmetry and inflation}},  \href{http://arxiv.org/abs/2003.08516}{{\tt
  arXiv:2003.08516}}.

\bibitem{Das:2020kff}
N.~Das and S.~Panda, {\it {Inflation in f(R,h) theory formulated in the
  Palatini formalism}},  \href{http://arxiv.org/abs/2005.14054}{{\tt
  arXiv:2005.14054}}.

\bibitem{Jarv:2020qqm}
L.~Järv, A.~Karam, A.~Kozak, A.~Lykkas, A.~Racioppi, and M.~Saal, {\it {On the
  equivalence of inflationary models between the metric and Palatini
  formulation of scalar-tensor theories}},
  \href{http://arxiv.org/abs/2005.14571}{{\tt arXiv:2005.14571}}.

\bibitem{Gialamas:2020snr}
I.~D. Gialamas, A.~Karam, and A.~Racioppi, {\it {Dynamically induced Planck
  scale and inflation in the Palatini formulation}},
  \href{http://arxiv.org/abs/2006.09124}{{\tt arXiv:2006.09124}}.

\bibitem{PhysRevD.42.2491}
J.~H. Traschen and R.~H. Brandenberger, {\it Particle production during
  out-of-equilibrium phase transitions},  {\em Phys. Rev. D} {\bf 42} (Oct,
  1990) 2491--2504.

\bibitem{Kofman:1994rk}
L.~Kofman, A.~D. Linde, and A.~A. Starobinsky, {\it {Reheating after
  inflation}},  {\em Phys. Rev. Lett.} {\bf 73} (1994) 3195--3198,
  [\href{http://arxiv.org/abs/hep-th/9405187}{{\tt hep-th/9405187}}].

\bibitem{Shtanov:1994ce}
Y.~Shtanov, J.~H. Traschen, and R.~H. Brandenberger, {\it {Universe reheating
  after inflation}},  {\em Phys. Rev. D} {\bf 51} (1995) 5438--5455,
  [\href{http://arxiv.org/abs/hep-ph/9407247}{{\tt hep-ph/9407247}}].

\bibitem{Kofman:1997yn}
L.~Kofman, A.~D. Linde, and A.~A. Starobinsky, {\it {Towards the theory of
  reheating after inflation}},  {\em Phys. Rev. D} {\bf 56} (1997) 3258--3295,
  [\href{http://arxiv.org/abs/hep-ph/9704452}{{\tt hep-ph/9704452}}].

\bibitem{Bezrukov:2008ut}
F.~Bezrukov, D.~Gorbunov, and M.~Shaposhnikov, {\it {On initial conditions for
  the Hot Big Bang}},  {\em JCAP} {\bf 06} (2009) 029,
  [\href{http://arxiv.org/abs/0812.3622}{{\tt arXiv:0812.3622}}].

\bibitem{GarciaBellido:2008ab}
J.~Garcia-Bellido, D.~G. Figueroa, and J.~Rubio, {\it {Preheating in the
  Standard Model with the Higgs-Inflaton coupled to gravity}},  {\em Phys. Rev.
  D} {\bf 79} (2009) 063531, [\href{http://arxiv.org/abs/0812.4624}{{\tt
  arXiv:0812.4624}}].

\bibitem{DeCross:2015uza}
M.~P. DeCross, D.~I. Kaiser, A.~Prabhu, C.~Prescod-Weinstein, and E.~I.
  Sfakianakis, {\it {Preheating after Multifield Inflation with Nonminimal
  Couplings, I: Covariant Formalism and Attractor Behavior}},  {\em Phys. Rev.
  D} {\bf 97} (2018), no.~2 023526,
  [\href{http://arxiv.org/abs/1510.08553}{{\tt arXiv:1510.08553}}].

\bibitem{Repond:2016sol}
J.~Repond and J.~Rubio, {\it {Combined Preheating on the lattice with
  applications to Higgs inflation}},  {\em JCAP} {\bf 07} (2016) 043,
  [\href{http://arxiv.org/abs/1604.08238}{{\tt arXiv:1604.08238}}].

\bibitem{Ema:2016dny}
Y.~Ema, R.~Jinno, K.~Mukaida, and K.~Nakayama, {\it {Violent Preheating in
  Inflation with Nonminimal Coupling}},  {\em JCAP} {\bf 02} (2017) 045,
  [\href{http://arxiv.org/abs/1609.05209}{{\tt arXiv:1609.05209}}].

\bibitem{DeCross:2016fdz}
M.~P. DeCross, D.~I. Kaiser, A.~Prabhu, C.~Prescod-Weinstein, and E.~I.
  Sfakianakis, {\it {Preheating after multifield inflation with nonminimal
  couplings, II: Resonance Structure}},  {\em Phys. Rev. D} {\bf 97} (2018),
  no.~2 023527, [\href{http://arxiv.org/abs/1610.08868}{{\tt
  arXiv:1610.08868}}].

\bibitem{DeCross:2016cbs}
M.~P. DeCross, D.~I. Kaiser, A.~Prabhu, C.~Prescod-Weinstein, and E.~I.
  Sfakianakis, {\it {Preheating after multifield inflation with nonminimal
  couplings, III: Dynamical spacetime results}},  {\em Phys. Rev. D} {\bf 97}
  (2018), no.~2 023528, [\href{http://arxiv.org/abs/1610.08916}{{\tt
  arXiv:1610.08916}}].

\bibitem{Sfakianakis:2018lzf}
E.~I. Sfakianakis and J.~van~de Vis, {\it {Preheating after Higgs Inflation:
  Self-Resonance and Gauge boson production}},  {\em Phys. Rev. D} {\bf 99}
  (2019), no.~8 083519, [\href{http://arxiv.org/abs/1810.01304}{{\tt
  arXiv:1810.01304}}].

\bibitem{Felder:2000hj}
G.~N. Felder, J.~Garcia-Bellido, P.~B. Greene, L.~Kofman, A.~D. Linde, and
  I.~Tkachev, {\it {Dynamics of symmetry breaking and tachyonic preheating}},
  {\em Phys. Rev. Lett.} {\bf 87} (2001) 011601,
  [\href{http://arxiv.org/abs/hep-ph/0012142}{{\tt hep-ph/0012142}}].

\bibitem{Felder:2001kt}
G.~N. Felder, L.~Kofman, and A.~D. Linde, {\it {Tachyonic instability and
  dynamics of spontaneous symmetry breaking}},  {\em Phys. Rev. D} {\bf 64}
  (2001) 123517, [\href{http://arxiv.org/abs/hep-th/0106179}{{\tt
  hep-th/0106179}}].

\bibitem{Akrami:2018odb}
{\bf Planck} Collaboration, Y.~Akrami et~al., {\it {Planck 2018 results. X.
  Constraints on inflation}},  \href{http://arxiv.org/abs/1807.06211}{{\tt
  arXiv:1807.06211}}.

\bibitem{Birrell:1982ix}
N.~Birrell and P.~Davies, {\em {Quantum Fields in Curved Space}}.
\newblock Cambridge Monographs on Mathematical Physics. Cambridge Univ. Press,
  Cambridge, UK, 2, 1984.

\bibitem{Chung:2004nh}
D.~J. Chung, E.~W. Kolb, A.~Riotto, and L.~Senatore, {\it {Isocurvature
  constraints on gravitationally produced superheavy dark matter}},  {\em Phys.
  Rev. D} {\bf 72} (2005) 023511,
  [\href{http://arxiv.org/abs/astro-ph/0411468}{{\tt astro-ph/0411468}}].

\bibitem{Tenkanen:2019aij}
T.~Tenkanen, {\it {Dark matter from scalar field fluctuations}},  {\em Phys.\
  Rev.\ Lett.} {\bf 123} (2019), no.~6 061302,
  [\href{http://arxiv.org/abs/1905.01214}{{\tt arXiv:1905.01214}}].

\bibitem{Fukugita:1986hr}
M.~Fukugita and T.~Yanagida, {\it {Baryogenesis Without Grand Unification}},
  {\em Phys. Lett. B} {\bf 174} (1986) 45--47.

\bibitem{Davidson:2008bu}
S.~Davidson, E.~Nardi, and Y.~Nir, {\it {Leptogenesis}},  {\em Phys. Rept.}
  {\bf 466} (2008) 105--177, [\href{http://arxiv.org/abs/0802.2962}{{\tt
  arXiv:0802.2962}}].

\bibitem{Lyth:2009zz}
D.~H. Lyth and A.~R. Liddle, {\em {The primordial density perturbation:
  Cosmology, inflation and the origin of structure}}.
\newblock 2009.

\bibitem{Peskin:1995ev}
M.~E. Peskin and D.~V. Schroeder, {\em {An Introduction to quantum field
  theory}}.
\newblock Addison-Wesley, Reading, USA, 1995.

\end{thebibliography}\endgroup

\end{document}